\documentclass[fleqn,12pt,twoside]{article}
\usepackage[headings]{espcrc1}

\usepackage{graphicx}
\usepackage[figuresright]{rotating}

\newcommand{\AmS}{{\protect\the\textfont2
  A\kern-.1667em\lower.5ex\hbox{M}\kern-.125emS}}

\def \ee {\end{eqnarray}} 
\def \be{\begin{eqnarray}&&}
\def \nonu {\nonumber \\&&}
\def\psla{\slash \! \! \!}

\begin{document} 
\title{Electromagnetic Hadron Form Factors 
and Higher Fock Components}

\author{J. P. B. C. de Melo 
\address{Centro de Ci\^encias Exatas e Tecnol\'ogicas,
 Universidade Cruzeiro do Sul, 08060-070,  and
Instituto de F\'\i sica Te\'orica, Universidade Estadual Paulista
 01405-900, S\~ao Paulo, Brazil}, 
 T. Frederico
\address{Dep. de F\'\i sica, Instituto Tecnol\'ogico da Aeron\'autica,
Centro T\'ecnico Aeroespacial, 12.228-900 S\~ao Jos\'e dos
Campos, S\~ao Paulo, Brazil},
 E. Pace and S. Pisano 
 \address{Dipartimento di Fisica, Universit\`{a} di
Roma "Tor Vergata" and Istituto Nazionale di \\ Fisica Nucleare, Sezione Tor
Vergata, Via della Ricerca Scientifica 1, I-00133 Roma, Italy}, 
 G. Salm\`e
\address{Istituto Nazionale di Fisica Nucleare, Sezione di Roma  \\ P.le A. Moro
2, I-00185 Roma, Italy} }


\maketitle

\begin{abstract} Investigation of the spacelike and timelike electromagnetic
form factors of hadrons, within a relativistic microscopical model characterized by
 a small set of
hypothesis, could shed light on the components of  hadron states beyond the valence one. Our relativistic 
approach  has been successfully  applied first to the pion and then 
 the extension to the nucleon has been undertaken. The pion  case is shortly 
 reviewed as an illustrative example
  for
introducing the main ingredients of our approach,
and preliminary results for the nucleon in the spacelike range $-10~ (GeV/c)^2\le q^2 \le 0$
 are evaluated.
\end{abstract}            

\section{INTRODUCTION}
The global analysis of the  electromagnetic (EM) form factors 
in both the spacelike (SL) region and  the timelike (TL)
one, within an approach based on the  Light-Front language \cite{dirac,brodsky},
 allows one to investigate in great detail the 
issue of non-valence components of both  hadron and  
photon wave functions. In Refs. \cite{prd06,DFPS}, the pion EM form factor,
in the range of momentum transfer $-10 ~(GeV/c)^2~\le q^2 \le 10
~(GeV/c)^2$, has been analyzed  within our relativistic approach, based on: i)  
a microscopic 
Vector Meson Dominance (VMD) model for the 
dressed vertex of the  photon,
and ii) a simple parametrization for the emission/absorption of a pion 
by a quark.  Indeed, the pion represents a suitable introduction for
illustrating the main features of our approach \cite{prd06,DFPS}.

Light-front  dynamics (LFD) approaches
\cite{brodsky}
 yield a unique possibility to study
the hadronic state, in both  the valence and the nonvalence sectors, since
within LFD a Fock series for the state of  a fermionic system, like 
\be
| meson \rangle =  ~|q\bar{q} \rangle + |q \bar{q} q \bar{q}\rangle +
|q \bar{q} ~g\rangle
 ..... 
\nonu
| baryon \rangle = |qqq \rangle + |qqq~q \bar{q} \rangle +|qqq~g \rangle
 ..... \nonumber 
\ee
 is a meaningful tool for investigating   hadronic properties. 
It is worth noting that the so-called zero
modes allow one to produce the breaking of the chiral symmetry through a
different mechanism from the standard one, so that   the LF vacuum acquires a
more simple structure.
Moreover, coupling the study of the EM form factors in the TL region with the
analysis in the SL one,  yields the
 possibility to
address the vast phenomenology of hadronic resonances, e.g.,  the  vector meson (VM)
propagation, and consequently  to impose
 strong constraints 
 on dynamical models devised for  a microscopical description of hadrons.
  Finally, 
 one can  obtain insight into  the two-body currents associated
to the $q \bar{q}$ 
pair production, that plays a very relevant role   in  a reference frame where 
$q^+ \ne 0 $ (see, e.g. \cite{LPS}), namely the frame mandatory for carrying on our global
analysis of the EM form factors.
\section{The Mandelstam formula for the EM current}
\indent  Our guidance for constructing a microscopical model that embeds,
 at large extent, the
advantages of a LF approach is given by 
the Mandelstam formula \cite{mandel} for the matrix elements of the EM current
of a system of interacting constituents. In a  global investigation 
of SL and TL regions one needs to change frame, from the $q^+=0$ frame (the 
standard choice within LFD, that always leads to $q^2\le 0$) to a 
$q^+\ne 0$ frame, that allows both positive and negative $q^2$. 
Therefore, a  covariant expression for the matrix elements of the 
EM current of  hadrons is a useful tool for constructing viable approximations.

 For instance, in the TL region the Mandelstam formula reads as follows
\be 
j^{\mu} = -\imath  e\int
\frac{d^4k}{(2\pi)^4}Tr\left [S_Q(k_Q) \bar{\lambda}_{h}(q-k,k_Q,P_{h})
 S(q-k)  \Gamma^\mu(k,q)
 S(-k) 
{\lambda}_{\bar h}(k,k_Q,P_{\bar h}) \right]    
\label{cur}\ee 
where
 $
S(p)=1/(\psla p-m+\imath \epsilon) $, 
with $m$ the mass of the constituent  struck by the virtual photon,
 $ S_Q(p)$ is the propagator of
 the spectator constituent (e.g. a quark, or a diquark in a simple
picture of baryons), $\lambda_{h}(k,k',P_{h})$  is the hadron vertex function,
which 
 contains the Dirac structure, i.e. a proper combination of Dirac
 matrices, and a momentum dependence;
 $P^{\mu}_{h}$ ($P^{\mu}_{\bar h}$) is the hadron (antihadron) momentum;
   $\Gamma^\mu(k,q)$ is the quark-photon vertex, and $q^{\mu}$ is the 
virtual photon momentum. The corresponding SL expression can be obtained by
applying the suitable changes to the hadron momenta and to the Dirac structure.
 The pion case is illustrated in detail in Refs. \cite{prd06,DFPS}. It should be
 pointed out that the momentum components of both $\lambda_{h}$  and 
 $\Gamma^\mu$ are  necessary for regularizing Eq. (\ref{cur}).

 Projecting out the 
Mandelstam formula on the Light-Front hyperplane,
through a $k^-$ integration, allows us to introduce sensible approximations for
constructing our phenomenological model. Let us remind that, presently, solving 
 the Bethe-Salpeter equation for fermionic system represents a big challenge,
 and therefore phenomenological studies could play a very useful role.
 
Our assumptions for evaluating the $k^-$ integration of Eq. (\ref{cur})
are: i) the momentum  components
of the vertex functions,
both for the pion and for the vector mesons entering in our VMD model for 
 $\Gamma^\mu(k,q)$,
vanish in the $k^-$ complex plane  for
$|k^-|\rightarrow\infty$,  ii) the contributions of the vertex function
singularities  can be neglected.  These approximations lead to the
following issues: i)
how to connect the Fock language with the
Bethe-Salpeter one, e.g. how to describe  both the amplitude for the
emission/absorption of a pion by a quark (i.e. the nonvalence
component of the pion wave function)  and the
$q\bar{q}$-pion vertex (i.e. the valence component), ii) how to model the
quark-photon vertex. In the limit $m_\pi \to 0$,
 only the contribution generated by 
 the nonvalence component of the hadron state, i.e. a higher Fock component, 
 is acting 
 in  both  TL and SL regions
 \cite{prd06,DFPS}, and  therefore the analysis is remarkably simplified. In particular,
in order to describe the emission/absorption of a pion
 by a quark, following Ref. \cite{JI01}
we assumed a pointlike interaction. It is worth
noting that in the present approach, the
coupling constant is fixed by the charge normalization of the pion form factor. 
 
 As to the quark-photon vertex, the bare term can be dropped, given our
 assumption on the pion mass, and one remains with the effect due to  the
$q\bar{q}$ production. In view of this, 
a  VMD approximation has been applied, namely the relevant
plus component of $\Gamma^\mu(k,q)$ is spanned over a VM basis as follows
  \be
 \Gamma^+(k,q) = \sqrt{2} \sum_{n, \lambda}~
\left [ \epsilon_{\lambda} \cdot \widehat{V}_{n}(k,k-P_n)  \right ]  ~ 
\Lambda_{n}(k,P_n) ~ 
{  [\epsilon ^{+}_{\lambda}]^* f_{Vn} \over (q^2 -
M^2_n + \imath M_n \tilde\Gamma_n(q^2))}  
\ee
where $f_{Vn}$ is the decay constant of the n-th vector
meson into a virtual photon ({\em to be calculated within our model}), 
 $M_n$  the corresponding mass, $\tilde\Gamma_n(q^2)$ the 
 total decay width (equal to   $\Gamma_n q^2/M^2_n$ for $q^2>0$  and
 vanishing for  $q^2\le0$) and
 $\epsilon_{\lambda}$  the VM polarization.   The  VM vertex amplitude  is 
 $\left [ \epsilon_{\lambda} \cdot \widehat{V}_{n}(k,k-P_n)  \right ]  ~ 
\Lambda_{n}(k,P_n)$, with a proper Dirac structure, $\widehat{V}_{n}(k,k-P_n)$, that   
generates the  Melosh rotations for $^3S_1$ states \cite{Jaus90} and  
 a momentum-dependent part indicated by $\Lambda_{n}(k,P_n)$. 
In the kinematical region corresponding to the valence
sector, the Bethe-Salpeter amplitudes, for both 
pion and  vector mesons,  are 
estimated through the eigenvectors of a relativistic simple model for
pseudoscalar and vector mesons \cite{FPZ02}.

An estimate of the probability of the valence component,
$P_{q\bar q}$, is necessary for a proper normalization of such a component 
for both
pion and vector mesons. In Ref. \cite{prd06} a simple and intuitive model for 
$P_{q\bar q}$ is presented.  

In the actual calculation,  20 vector mesons are taken
into account to reach  convergence up to $|q^2|=10~(GeV/c)^2$. 
 The corresponding eigenstates are calculated from
 the  model of Ref. \cite{FPZ02}, while for  the masses a different choice has
 been made: for the first four vector mesons, the experimental values (together
 with the corresponding widths) have been considered, while for $M_{n} >2.150~ GeV$ 
 the model eigenvalues have
 been adopted.   We have introduced {\em two adjusted parameters}:
i) a single width, $\Gamma_n$,  for the vector mesons with mass 
$M_{n}>~ 2.150~ GeV$, ii)  the  relative weight, $w_{VM}$, of the so-called 
"instantaneous" contributions of the  pion and the vector mesons. These instantaneous
terms are generated by the instantaneous
part of the quark propagator, and they  play the central role in the 
evaluation of
the   pion form factor,  in the limit of a vanishing pion mass.
We have chosen for the experimentally unknown widths 
a value  $\Gamma_n=0.15~GeV$,   similar to the experimental width of the first 
four VM's; 
while for $w_{VM}$
we have found that $w_{VM}=-0.7$ is a good value for  a global fit 
(see Fig. 1), and   $w_{VM}=-1.5$ gives
 an improved
description of the $\rho$ peak region (see Figs. 10 and 11 of \cite{prd06}).

\begin{figure}[t]
\parbox{10.2cm}{
\includegraphics[width=10.1cm]{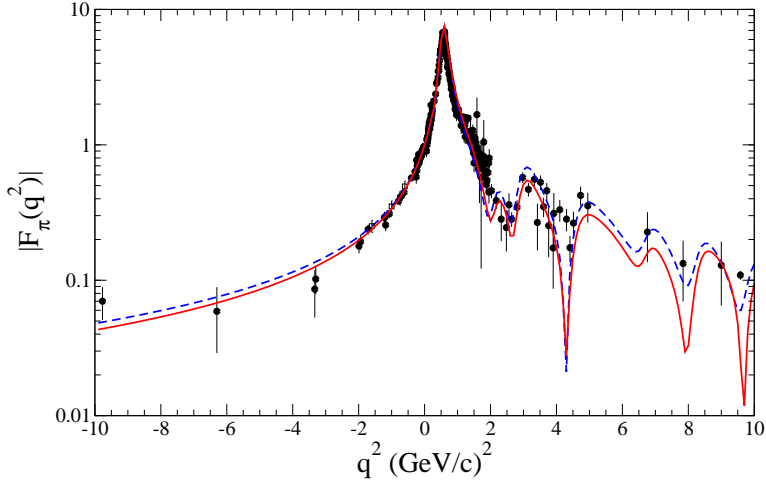}}
$~$
\parbox{52mm}{
\caption{Pion EM form factor in the SL and TL regions vs $q^2$. 
Solid line: calculations with  the pion wave function  from the model of Ref.
\cite{FPZ02}, and 
a relative weight $w_{VM}=-0.7$ for the "instantaneous" terms. Dashed line: 
the same as the solid line, but with the asymptotic pion 
wave function \cite{lepag}. Data from the collection of Ref.
\cite{baldini}. (After Ref. \cite{prd06})}
      }   

\end{figure}

 \section{The EM nucleon form factors}
The evaluation of the Sachs form factors of the nucleon largely proceeds along
the same line of the pion. We consider the nucleon as a system of quarks with 
mass equal to $.220~
GeV$, and then we model the quark-nucleon amplitude in analogy with the pion case.
First of all, we adopt a Dirac structure of 
the quark-nucleon vertex as suggested by an effective Lagrangian density like the one of Ref.
\cite{wilson}, viz
\be
{\cal L}_{eff}(x,\tau_1,\tau_2,\tau_3,\tau_N) =  \epsilon _{abc} ~\int d^4x_1~d^4x_2~d^4x_3
{\cal F}(x_1,x_2,x_3,x)~\times \nonu
\left [m_N~\alpha~{1 \over \sqrt{2}}\sum_{\tau_1,\tau_2,\tau_3}
~\bar q^a(x_1){\cal T}_{\tau_1}^\dagger  \imath \tau_y~{\cal T}_{\tau_2}^*~
\gamma^5
~q^b_C(x_2)~\bar q^c(x_3) {\cal T}_{\tau_3}^\dagger~+ \right. \nonu \left.
-~(1-\alpha)~{1 \over \sqrt{6}}~ \sum_{\tau_1,\tau_2,\tau_3}
~
 \bar q ^a(x_1) {\cal T}_{\tau_1}^\dagger ~\vec \tau~\imath \tau_y~{\cal T}_{\tau_2}^*~
\gamma^5~\gamma_\mu~
q^b_C(x_2)\cdot\bar q^c(x_3) {\cal T}_{\tau_3}^\dagger \vec \tau~(-\imath~\partial^\mu)\right ]
~\psi_N(x){\cal T}_{\tau_N}+
\nonu
+cyclic + h.c.
\label{lag}\ee
where ${\cal F}(x_1,x_2,x_3,x)$ is a scalar distribution that properly weights
  the quark coordinates; $\epsilon _{abc}$ takes care of the color degrees of freedom; $m_N$ is the
mass of the nucleon; $q(x,\tau)=q(x)~{\cal T}_\tau$ is the field that creates a quark $u$ or $d$
depending upon $\tau$; the charge conjugated  
field is 
$q_C(x,\tau)={\cal C}q(x,\tau){\cal C}^\dagger= C\bar q^T(x)~
{\cal T}_\tau^C=C\bar q^T(x)~
\imath~\tau_y~{\cal T}_\tau^*=
\imath\gamma^2\gamma^0 \bar
q^T(x)\imath~\tau_y {\cal T}_\tau^*$ and  has a parity
opposite to the quark field (this entails 
  a pseudoscalar  or an axial operator between  $\bar q^a$ and
$q_C^b$, in order to recover the correct intrinsic parity of the nucleon). The
isospin field ${\cal T}_\tau^C=\imath~\tau_y~{\cal T}_\tau^*$ 
 creates a charge conjugate isospin state, namely 
 $\imath~\tau_y~\chi^*_{\tau_2}$.
For the present time $\alpha=1$,  i.e. we have considered a pseudoscalar coupling only. 

In what follows, given the preliminary character of this presentation, we  deal with the 
discussion of the EM form factor in the SL
region. The matrix elements of the {\em macroscopic} EM current of the nucleon,
\be \langle  \sigma',p'|j^\mu~|p,\sigma\rangle= \bar U_N(p',\sigma')\left [-F_2(Q^2) { {p'}^\mu +{p}^\mu \over 2M_N}
+\left (F_1(Q^2)+F_2(Q^2)\right )~\gamma^\mu\right] U_N(p,\sigma)
\label{macro}\ee
where $Q^2=-q^2$ and $F_{1(2)}$ is the Dirac (Pauli) nucleon form factor, are approximated {\em microscopically} as follows
\be
\langle  \sigma',p'|j^\mu~|p,\sigma\rangle
=~N_c~\int {d^4k_1 \over (2\pi)^4}\int {d^4k_2 \over (2\pi)^4} \times \nonu
 Tr
\left \{\bar \Phi^{\sigma'}_N(k_1,k_2,k'_3,p')~S^{-1}(k_1)~S^{-1}(k_2)~
{\cal I}^\mu_3~
 ~\Phi^\sigma_N(k_1,k_2,k_3,p)\right \}
\label{spc1}\ee
where the factor $N_c$ is a color weight, the trace is performed over isospin and Dirac matrices, 
$\Phi^\sigma_N(k_1,k_2,k_3,p) $ is the quark-nucleon Bethe-Salpeter amplitude \cite{nucl06}, that contains both a Dirac
structure as suggested by Eq. (\ref{lag}) and a momentum component, as in  the pion case;
${\cal I}^\mu$ is the quark-photon vertex with isoscalar 
and isovector
   contributions, namely 
  \be
  {\cal I}^\mu=~\left ({1 +\tau_z \over 2} \right )~{\cal I}^\mu_u 
  +\left ({1 -\tau_z \over 2} \right )~{\cal I}^\mu_d=
  {{\cal I}^\mu_u+{\cal I}^\mu_d \over 2}+\tau_z{{\cal I}^\mu_u-{\cal I}^\mu_d 
  \over 2}={\cal I}^\mu_{IS} +\tau_z
  {\cal I}^\mu_{IV}
 \label{cur1} \ee
  The terms ${\cal I}^\mu_{IS}$ and $
  {\cal I}^\mu_{IV}$ contain a  pointlike (bare) contribution and a VMD contribution,
   with isoscalar 
  ($\omega$-like) and isovector ($\rho$-like) content, respectively,
 viz
\be
{\cal I}^\mu_{IS(IV)}(k,q) = {\cal N}_{IS(IV)}\theta(p^+-k^+)~\theta(k^+)
  \gamma^\mu~ +
  \theta({p'}^+ - k^+)~
 \theta(k^+-{p}^+)~\times \nonu\left \{{Z_b}~{\cal N}_{IS(IV)}\gamma^\mu+ 
   {Z_{V}~}\Gamma^\mu_{V}[k,q,IS(IV)]\right\}
 \label{vert}\ee
 where ${\cal N}_{IS}=1/6$ and ${\cal N}_{IS}=1/2$,   the quantities
 ${Z_b}$ (bare term) and ${Z_{V}}$ (VMD term) are unknown renormalization constants to be extracted from
 the phenomenological analysis of the data. The theta functions, acting on the plus component  of the
 momenta involved in the process, implement the kinematical constraints obtained from the 
 projection of Eq. (\ref{spc1}) on the LF hyperplane, as in the case of the pion. In particular,
 they identify the kinematical region relevant to the physical processes 
 depicted in Fig. 2, that can be interpreted as valence and nonvalence
contributions. Moreover, the vertical dashed lines in Fig. 2 allow one to count
 how many constituents are
in flight, at a given LF time. It is worth noting that while the VMD is uniquely associated
to the pair production by the virtual photon,  (cf  contribution $(b)$ in  Fig. 2), 
i.e. to the nonvalence component of the nucleon, the bare term ($\gamma^\mu$) 
can yield a contribution
  both in the elastic quark-photon
channel (or triangle diagram, cf  contribution $(a)$ in  Fig. 2) and  in 
the pair production process.

The momentum dependence of the quark-nucleon
amplitude in the valence sector,   where the spectator quarks are on their-own
$k^-$-shell, has been modeled  adopting a power-law Ansatz \cite{lepag}, like
\be
{\cal W}_N(k_1,k_3,k_3,p)\sim {1 \over \left[\beta^2 +M^2_0(1,2,3)\right]^3}
\ee
where $\beta $ has been fixed through  the anomalous magnetic moments of the nucleon. In particular,
with 
$\beta =0.118 ~GeV$, one  obtains  $\mu_p= 2.878$ (Exp. $2.793$)  and $\mu_n=-1.859$ 
(Exp. $-1.913$), and $M_0(1,2,3)$ is the LF free mass for a three-quark system. 
In the non-valence sector, relevant for evaluating the Z-diagram
contribution, the momentum dependence  of the quark-nucleon
amplitude is approximated by
\be
G_N(k_1,k_3,k_3,p)\sim {1 \over \left [\beta^2+M^2_0(1,2)\right]^2}~\left \{ {1 \over \left
[\beta^2+M^2_0(3',2)\right]} +{1 \over \left
[\beta^2+M^2_0(3',1)\right]}\right \}
\ee
where $M_0(i,j)$ is the LF free mass for a two-quark system. This Ansatz is 
different from the one used for the pion
case, where a simple constant vertex was adopted, since the sixth-fold 
integration needs a 
stronger regularization.

For the present, preliminary calculations in the SL region we have the following
 adjusted parameters:
i) $Z_b$ and $Z_V$, that are  renormalization constants for the pair production terms in the
quark-photon vertex (cf Eq. (\ref{vert})), ii)
two parameters for a running mass in the valence sector (with a range of  variation  
of about 10$\%$, for $-10~ GeV^2~< q^2 <~0$),
 iii)
two parameters for a running mass in the non-valence sector (with a range of  variation  
of about 20$\%$, for $-10~ GeV^2~< q^2 <~0$).

 \begin{figure}[t]

\includegraphics[width=13cm]{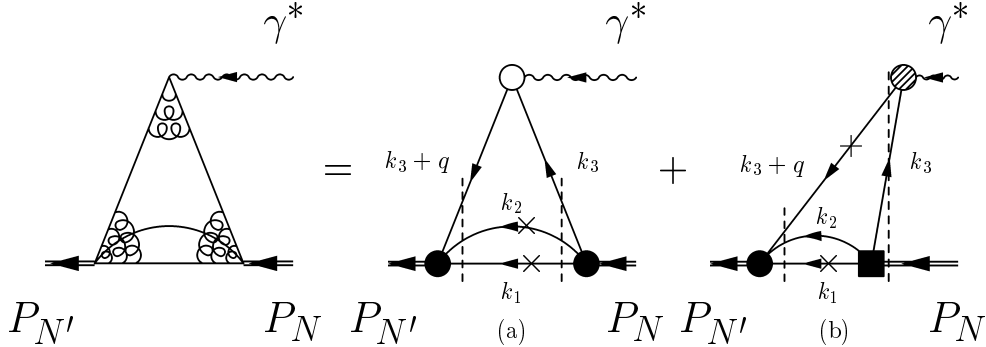} 

\caption{ Diagrams contributing to the nucleon EM form factors in the 
spacelike region. Diagram $(a)$ (triangle or elastic term) illustrates the valence sector contribution, where 
$0 < k_{i}^+ < P^+_{N}$ ($i=1,2,3$), and   the symbol $\times$ on a quark line indicates
that the corresponding quark is  on its $k^-$-shell. Diagram $(b)$ shows the nonvalence sector 
contribution,
$P^+_{N} < k_3^+ < {P'}^+_N$, due to the pair production. }
\end{figure}
 \begin{figure}
\parbox{7.8cm}{\includegraphics[width=7.8cm]{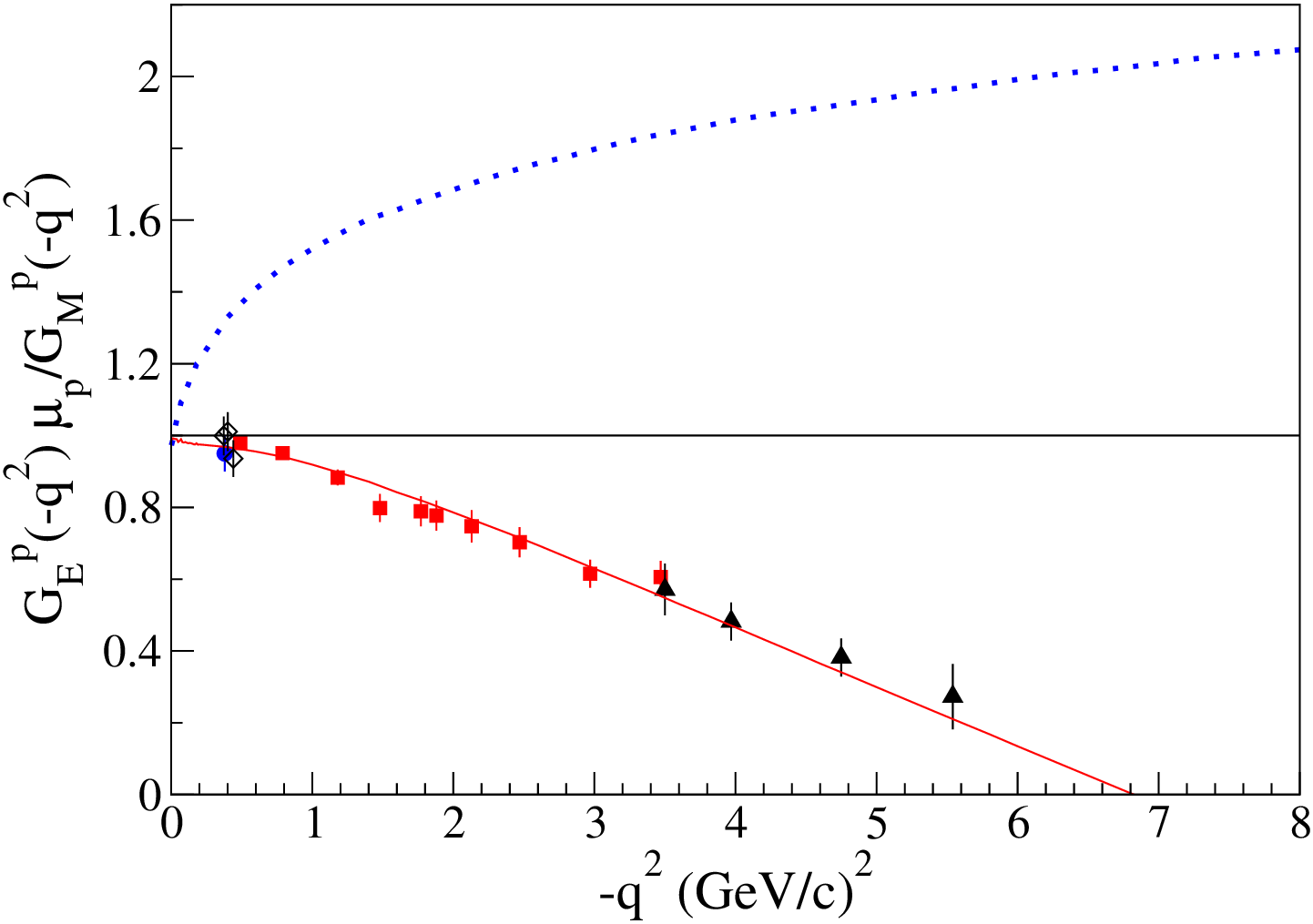}}
$~~$\parbox{7.8cm}{\includegraphics[width=7.8cm]{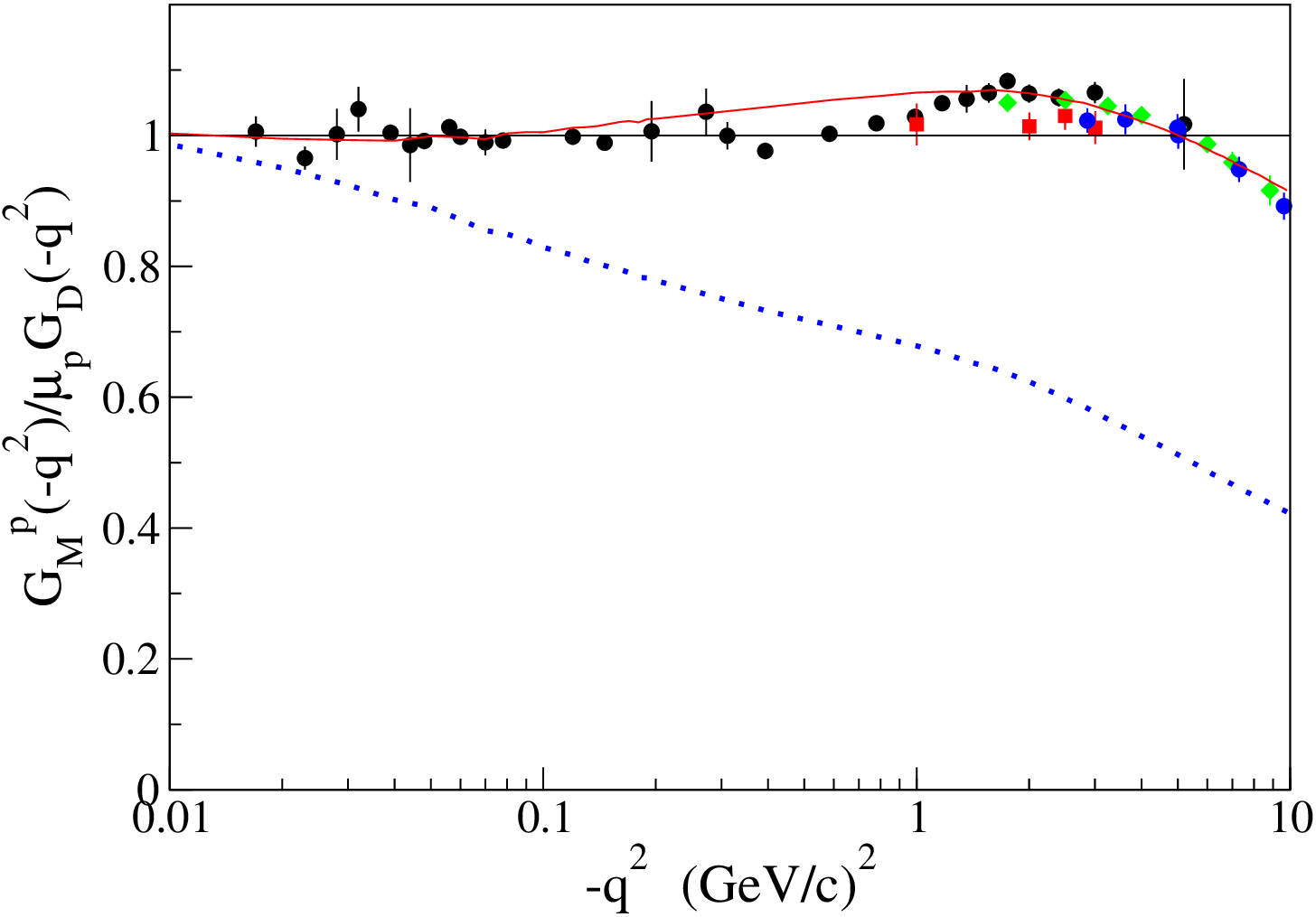}}
\caption{Left panel. The ratio  $G^p_E(Q^2) \mu_p/G^p_M(Q^2)$ vs $Q^2$.
Solid line: full calculation, corresponding to the sum  of all the
 contributions to 
$G^p_E(Q^2) $ and $G^p_M(Q^2)$, i.e. triangle plus pair production terms. 
Dotted line: triangle elastic contribution to $G^p_E(Q^2) $ and $G^p_M(Q^2)$.  
Data: from \cite{jlab}.
Right panel. $G^p_M(Q^2)/\mu_p G_D(Q^2)$ vs $Q^2$, with  the same notations as 
in the left panel, and  $ G_D(Q^2)=1/(1+Q^2/0.71)^2$.}

\vspace {1cm}
\parbox{7.8cm}{\includegraphics[width=7.8cm]{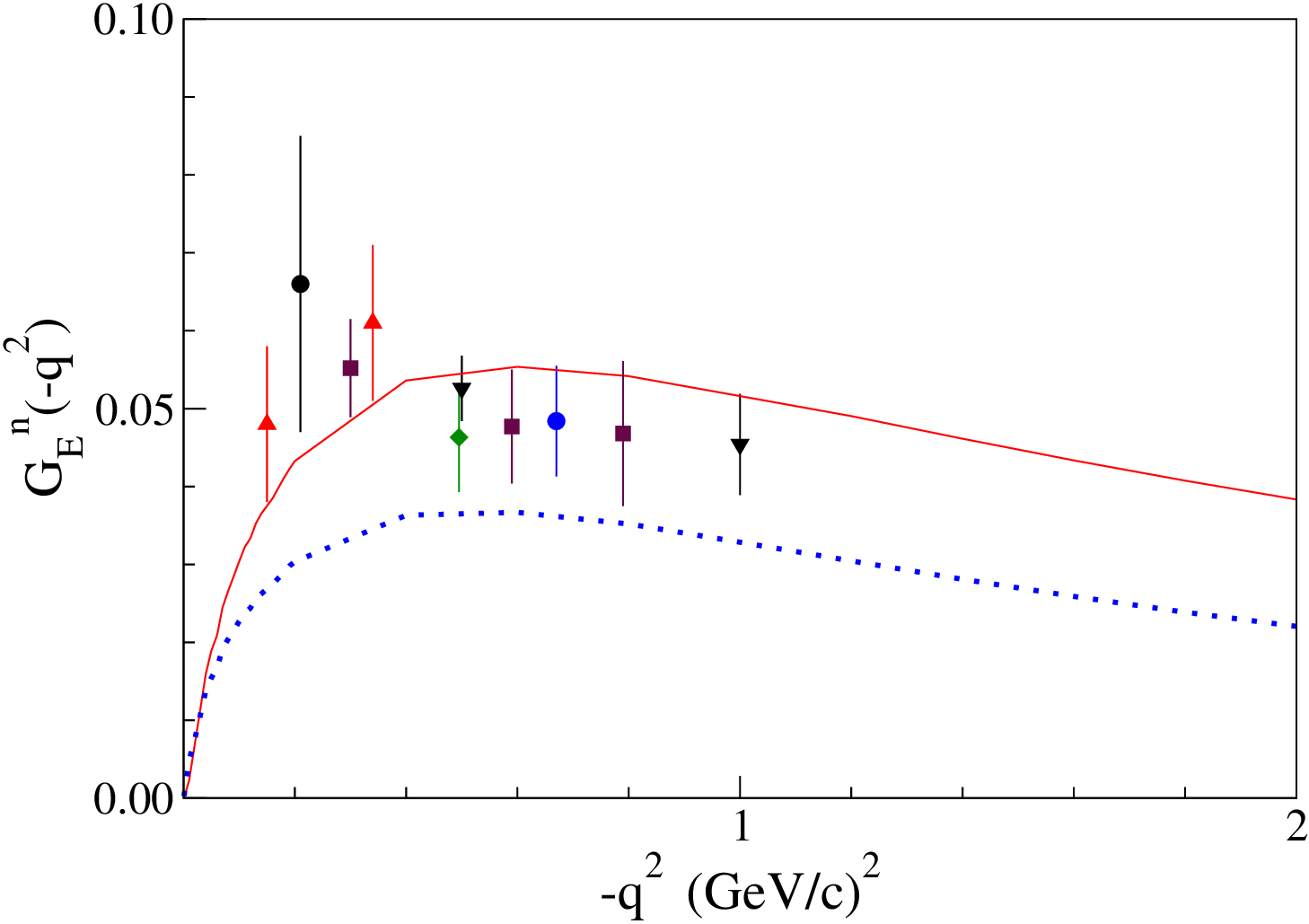}}
$~~$\parbox{7.8cm}{\includegraphics[width=7.8cm]{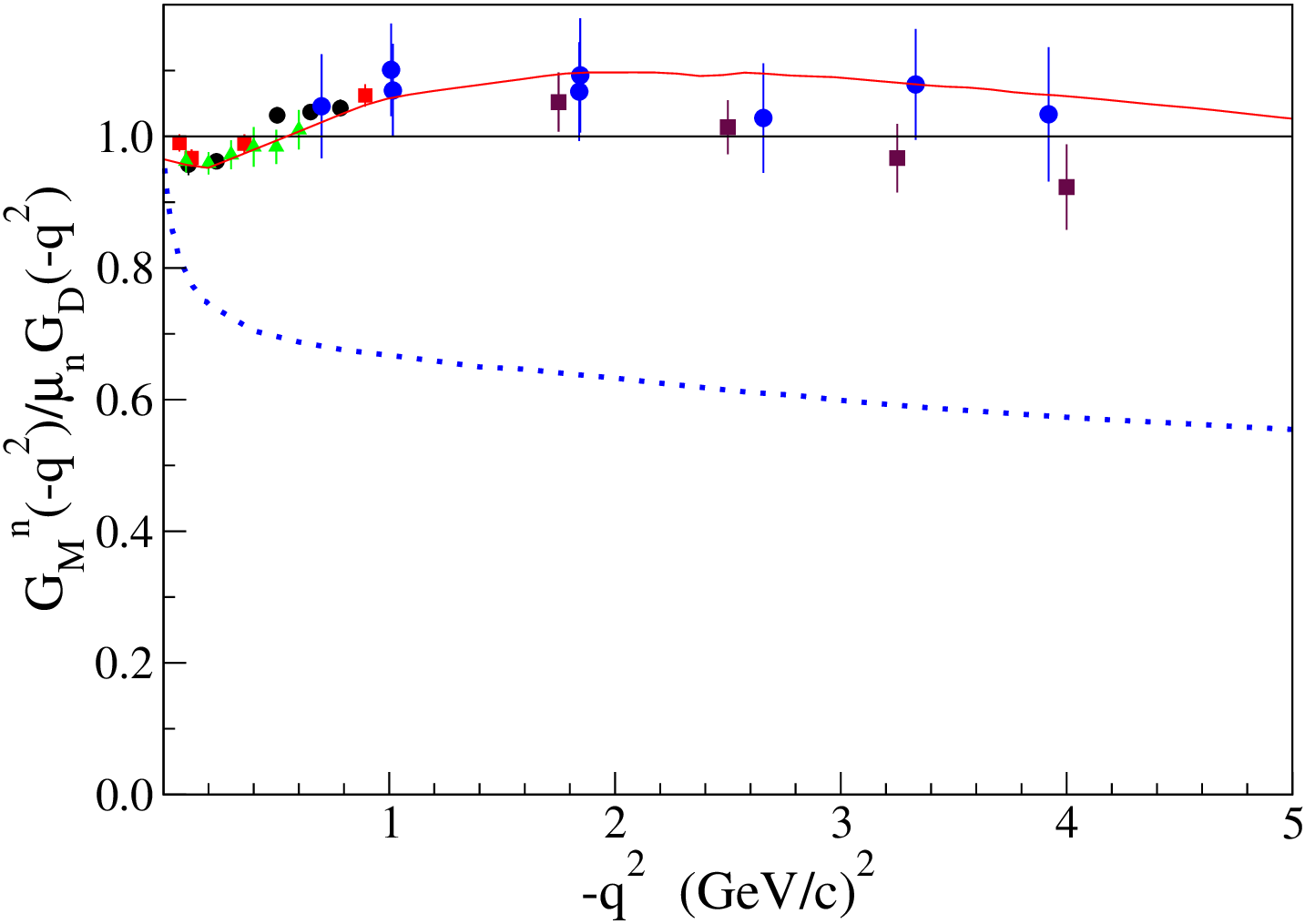}}
\caption{Left panel.   $G^n_E(Q^2) $ vs $Q^2$.
Solid line: full calculation, corresponding to the sum of all the contributions to 
$G^n_E(Q^2) $, i.e. triangle plus pair production terms. 
Dotted line: triangle elastic contribution to $G^n_E(Q^2) $.  
Data: from \cite{jlab}.
Right panel. $G^n_M(Q^2)/\mu_n G_D(Q^2)$ vs $Q^2$, with the same notations 
as in the left panel, and  $ G_D(Q^2)=1/(1+Q^2/0.71)^2$.}      
\end{figure}

The Sachs form factors can be obtained from the {\em macroscopic} current by
applying the  proper traces to Eq. (\ref{macro}) and then by using the 
microscopical evaluation 
of the
relevant matrix elements. In particular in the frame where
${\bf p}'_{\perp}={\bf p}_\perp=0$ and $q^+=\sqrt{Q^2}$ 
\be
G_E^N(Q^2) = F_1(Q^2)-{Q^2 \over 4 M^2_N} F_2(Q^2)=
\nonu ={  M_N^2\over 2 p^+ {p'}^+} Tr \left \{{\psla p'+M_N \over 2 M_N}~\left [-F^N_2(Q^2) { {p'}^+ +
{p}^+ \over 2M_N}
+\left (F^N_1(Q^2)+F^N_2(Q^2)\right )~\gamma^+\right]{\psla p+M_N \over 2 M_N} \gamma^+\right\}
\nonu
\nonu
G_M^N(Q^2) = F_1(Q^2)+ F_2(Q^2)=
\nonu ={ 2 M_N^2\over Q^2}~ Tr \left \{{\psla p'+M_N \over 2 M_N}~\left [-F^N_2(Q^2) { {p'}^1 +{p}^1 \over 2M_N}
+\left (F^N_1(Q^2)+F^N_2(Q^2)\right )\gamma^1\right] {\psla p+M_N \over 2 M_N}\gamma^1\right\}
\ee
In Figs. 3 and 4, the ratio $G^p_E(Q^2) \mu_p/G^p_M(Q^2)$, and the form factors
 $G^p_M(Q^2)$,
$G^n_E(Q^2)$, $G^n_M(Q^2)$ are shown. 
One of the main feature of our preliminary calculation is the explanation of 
the possible zero in 
the ratio $G^p_E(Q^2)
 \mu_p/G^p_M(Q^2)$ in terms of the interplay between the Z-diagram contribution, i.e.
higher Fock components, and the elastic contribution. Such an interpretation 
appears supported by the encouraging agreement between theoretical calculations
 and  experimental data. 

\section{Conclusions \& Perspectives}
In \cite{prd06,DFPS}, we constructed a microscopical model for  the
pion form factor, within a LF approach, obtaining a very nice description 
of experimental data. The development of this model for calculating EM
  nucleon form factors
 in both SL and TL regions is our next goal.
  Our model for the nucleon exploits many ingredients already introduced
 for the pion, adding suitable Ansatzes for the quark-nucleon amplitude in both the valence and
 nonvalence sectors. The possibility to address separately valence and
 nonvalence sectors of the hadron amplitude represents a distinct feature 
 of our phenomenological
  model, and it is entailed by  the 
  model for the quark-photon vertex, we have adopted. In particular,  within 
  our approach one could ascribe the possible
zero in the ratio $G^p_E(Q^2)
 \mu_p/G^p_M(Q^2)$  to the interplay between the contribution from the pair production process 
 and the one from the elastic
 quark-photon vertex, shedding some light on the deep meaning of the puzzling 
 experimental finding.

 The next step will be the extension of our  calculation to the timelike region.

\end{document}